\documentclass[a4,11pt]{article}
\usepackage{amssymb,amsthm,amscd, amsbsy, array}
 \usepackage{amsmath,bm,cite}
 \usepackage{yfonts}
   \usepackage{dsfont}
\usepackage{graphicx}
\newcommand{\mathsym}[1]{{}}

\usepackage{epsfig}
\usepackage{epsf}
\usepackage{rotating}

\let\ssection=\section
\renewcommand{\section}{\setcounter{equation}{0}\ssection}
\newcommand{\half}{{\scriptstyle{\frac{1}{2}}}}
\newfont{\tenmsb}{msbm10 scaled\magstep1}

\newcommand{\nn}{{\nonumber}}

\newcommand{\lf}{\left (}
\newcommand{\lfq}{\left [}

\newcommand{\rg}{\right )}
\newcommand{\rgq}{\right ]}

\def\smallover#1/#2{\hbox{$\textstyle{#1\over#2}$}}
\def\beq{\begin{equation}}
\def\eeq{\end{equation}}
\def\beqa{\begin{eqnarray}}
\def\eeqa{\end{eqnarray}}
\newcommand{\bean}{\begin{eqnarray*}}
\newcommand{\eean}{\end{eqnarray*}}
\def\lf{\left(}
\def\rg{\right)}
\def\lq{\left[}
\def\rq{\right]}
\def\lgr{\left\{}
\def\rgr{\right\}}
\def\p{{\partial}}
\def\p{{\partial}}

\def\*{{\star}}

\begin{document}

\title{Lie-point symmetries of the discrete Liouville equation}
\author{
D. Levi $^1$\footnote{e-mail:
Levi@roma3.infn.it}, L. Martina$^{2,3}$ \footnote{e-mail:
Luigi.Martina@le.infn.it} , P. Winternitz $^{1,4}$\footnote{e-mail:
wintern@crm.umontreal.ca}
\\
$^1$ Dipartimento di Matematica e Fisica, Universita' degli Studi Roma Tre, \\e Sezione INFN di Roma Tre,\\ Via della Vasca Navale 84, 00146 Roma (Italy)\\
$^{2 }$  Dipartimento di Matematica e Fisica - Universit\`a del Salento
\\
$^{3}$ Sezione INFN di Lecce. Via Arnesano, CP. 193 
I-73 100 Lecce (Italy) \\
$^4$ Centre des Recherches Mathematiques, Universit\'e de Montr\'eal, Montr\'eal (Qu\'ebec, Canada)
}

\maketitle

%%%%%%%%%%%
\begin{abstract}
The Liouville equation is well known to be linearizable by  a point transformation. It has an infinite dimensional Lie point symmetry algebra isomorphic to a direct sum of two Virasoro algebras. We show that it is not possible to discretize the equation keeping the entire symmetry algebra as point symmetries. We do however construct a difference system approximating the Liouville equation that is invariant under the maximal finite subalgebra $ SL_x \lf 2 , \mathbb{R} \rg \otimes SL_y \lf 2 , \mathbb{R} \rg $. The invariant scheme is an explicit one and provides a much better approximation
of exact solutions than comparable standard (non invariant) schemes.
\end{abstract}
%%%%%%%%%%%
\vspace{5mm}
\noindent
%%%%%%%%%%%%%%%%%%%%%%
\section{Introduction}
The purpose of this article is to investigate the possibility of discretizing the Liouville equation 
\beq z_{x y} = e^z \label{LiouvilleEq},\eeq
or its algebraic version 
\beq u\, u_{x y} -u_x\, u_y = u^3 , \qquad u = e^z, \label{AlgLiouvilleEq}\eeq
while preserving all of its Lie point symmetries. This is quite a challenge, since the Lie point symmetry group of these equations is infinite dimensional. We shall call (\ref{AlgLiouvilleEq}) the {\it algebraic Liouville equation}.

The article  is part of a general program on the  study  of continuous symmetries of discrete equations \cite {d91,Doro-book,dkw00,bcw06,brw08,LW2006,LOTW,LWY2002,lstw,ltw,LW1996,LW91,rv13,RebeloValiquette,rw09,W2004,WCUP2011}. This program has several aspects each possibly requiring different approaches. They are: 
\begin{enumerate}
\item In relativistic and nonrelativistic quantum mechanics or field theory on a discrete space--time, a  problem is to discretize the continuous theory while preserving continuous symmetries such as  rotational, Lorentz, Galilei or conformal invariance. One possible way of doing this is the way explored in the present article, namely to not  use a preconceived constant lattice. Instead one can   construct an invariant set of equations defining both the  lattice and system of difference equations. The lattice thus appears as part of a solution of a set of discrete equations and the symmetry group  acts on the solutions of the equation and on the lattice.  
\item The study of symmetries of genuinely discrete phenomena, such as molecular or atomic chains, where the discrete lattice is given a priori.
\item The third aspect  of this program fits into the general field of  geometrical integration \cite{mq06,mw01,i08,hlw06}. The basic idea is to improve numerical methods of solving specific ordinary and partial differential equations, by incorporating important qualitative features of these equations into their discretization. Such features may be integrability, linearizability, Lagrangian or Hamiltonian formulation, or some other features. 
\end{enumerate}
 We concentrate on the preservation of Lie point symmetries. In our case the idea is to take an ordinary or partial differential equation  (ODE or PDE) with a known Lie point symmetry algebra $\cal L$  realized by vector-fields. The differential equation is then approximated by a difference system with the same symmetry algebra. The difference system consists of a set of difference equations, describing both the approximation of the ODE (PDE) and the lattice. The difference system is constructed out of the invariants of the Lie point symmetry group $\cal G$ of the original ODE (PDE). The Lie algebra $\cal L$ of $\cal G$ is realized by the same vector fields as for the continuous  equation, however its action is prolonged to all points of the lattice, rather than to derivatives.  
 
 In Section 2 we present the Lie point symmetry algebra of the
continuous algebraic Liouville equation and the corresponding vector
fields depending on two arbitrary functions of one variable
each. The symmetry algebra is isomorphic to the direct sum of two Virasoro
algebras (with no central
extension).  We also give the two second order invariants of the  maximal
finite-dimensional subgroup $ SL_x \lf 2 , \mathbb{R} \rg \otimes SL_y \lf 2 , \mathbb{R} \rg $ of the corresponding infinite
dimensional symmetry group.
Section 3 is devoted to a brief exposition of the method of discretizing
differential equations
while preserving their point symmetries. In Section 4 we  discretize the
Liouville equation on a four-point stencil. The discretization is
invariant under the maximal finite dimensional subgroup, not however under
the entire infinite -dimensional group. Section 5 is devoted to numerical
experiments. We choose 3 different exact solutions of the continuous
Liouville equation and then formulate a boundary value problem that leads
to these solutions. The boundary value problem is then solved numerically,
using a standard discretization and our invariant one.
The results are compared to the exact solutions. In all three cases the
invariant discretization is shown to perform considerably better than the
standard one.  An alternative symmetry preserving discretization of the
Liouville equation due to Rebelo and Valiquette [23] is discussed in
Section 6. They have succeeded in preserving the entire symmetry group but
as generalized symmetries rather than point ones (only translations and
dilations remain as point symmetries). Finally, in Section 7 we discuss a
linearizable discretization due to Adler and Startsev [1] and show that it
has no continuous Lie point symmetries at all. The last Section 8 is
devoted to conclusions.
%%%%%%%%%%%%%%%%%%%%%%
\section{Lie point symmetries of the continuous Liouville equation}
The Liouville system (\ref{LiouvilleEq}) is a remarkable equation that has already been thoroughly investigated. It was shown by Liouville himself \cite{Liouville} that it is linearized into the linear wave equation by the transformation 
\beq z = \ln \lfq 2 \frac{\phi_x \, \phi_y}{\phi^2}\rgq, \qquad \phi_{x \, y} = 0 .\label{GenSol} \eeq
Putting $\phi\lf x, y \rg = \phi_1\lf x \rg + \phi_2\lf y \rg$, where $\phi_i, \; i=1, 2$ are arbitrary functions, we get a very general class of solutions of (\ref{LiouvilleEq}) ( and  (\ref{AlgLiouvilleEq}) ), namely
\beq z = \ln \lfq 2 \frac{\phi_{1,x} \, \phi_{2, y}}{\lf \phi_1 + \phi_2\rg^2 }\rgq.\label{GenSol2} \eeq

In view of (\ref{GenSol})  the Liouville equation is linearizable and it is not surprising that its symmetry algebra is infinite dimensional, as  was already known in 1898 \cite{Medolaghi}.  The symmetry algebra of the algebraic Liouville equation (\ref{AlgLiouvilleEq}) is given by the vector fields 
\beqa X\lf f\lf x \rg\rg = f\lf x \rg \p_x - f_x\lf x \rg \, u \, \p_u, \qquad 
Y\lf g\lf  y \rg\rg = g\lf  y \rg \p_y - g_y\lf  y \rg \, u \,  \p_u \label{genAlg},\eeqa
where $f = f\lf x \rg$ and $g = g\lf y \rg$ are arbitrary smooth functions. The nonzero commutation relations of the vector fields (\ref{genAlg}) are 
\beq \lq  X\lf f \rg , X ( \tilde f ) \rq   = X\lf f \tilde{f}_x- \tilde{f} f_x \rg , \quad   \lq  Y\lf g \rg , Y\lf \tilde{g}\rg \rq   = Y\lf g\, \tilde{g}_y - \tilde{g} \, g_y \rg , \quad \lq  X\lf f\rg , Y\lf {g}\rg \rq   = 0. \label{commrel}\eeq
The algebra (\ref{genAlg})-(\ref{commrel}) is isomorphic to the direct sum of two Virasoro algebras. We denote it $L = vir_x \oplus vir_y$. Its maximal finite dimensional subalgebra is  $ sl_x \lf 2 , \mathbb{R} \rg \bigoplus sl_y \lf 2 , \mathbb{R} \rg $, obtained by restricting $f\lf x \rg$ and $g 
\lf y \rg$ to be second order polynomials. Limiting ourselves to a neighborhood of the origin, the above vector fields can be expanded in the basis  $\lgr X\lf x^n \rg\rgr_{n\in \mathbb{N}}$ and $\lgr Y\lf y^n \rg\rgr_{n\in \mathbb{N}}$, which leads to the  commutation relations 
\beqa \lq X\lf x^m \rg, X\lf x^n \rg \rq &=& \lf n - m\rg X\lf x^{m+n-1} \rg , \quad \lq Y\lf y^m \rg, Y\lf y^n \rg \rq = \lf n - m\rg Y\lf y^{m+n-1}\rg , \nn \\  \quad \lq X\lf x^m \rg, Y\lf y^n \rg \rq &=& 0. \label{commrelpol}\eeqa
As said above, the maximal finite subalgebra corresponds to the basis elements with $m, n = 0,1,2$.

Let us find the most general second order expression of the form $I\lf x,y, u, u_x, u_y, u_{x x}, u_{x y}, u_{y y} \rg$ invariant under the group corresponding to the algebra (\ref{genAlg}). The second order prolongation of $X\lf f \rg$ is 
\beqa \textrm{pr}^{\lf 2 \rg}\, X\lf f\rg  &=& f \p_x - f' \lfq u\, \p_u + 2 u_x \, \p_{u_x} + u_y \, \p_{u_y}  
+ 2 u_{x y} \, \p_{u_{x  y}} + 3 u_{x  x} \, \p_{u_{x x}} +  u_{y y} \, \p_{u_{y y}} \rgq \nn \\
 &&- f'' \lfq u\, \p_{u_x} + u_y \p_{u_{x y}} + 3 u_x \p_{u_{x x}}\rgq - f''' u \p_{u_{x x}} \label{prolalg}\eeqa
 and similarly for $Y\lf g \rg$. We see  that the last term in (\ref{prolalg}) is absent in the subalgebra. 

The group $ SL_x \lf 2 , \mathbb{R} \rg \otimes SL_y \lf 2 , \mathbb{R} \rg $ allows  two functionally independent "strong" invariants, namely
\beq  I_1 = \frac{u u_{x y} - u_x\, u_y}{u^3} , \qquad I_2 = \frac{\lf 2 u u_{x x} - 3 u_x^2\rg \, \lf 2 u u_{y y} - 3 u_y^2\rg }{u^6}.  \label{InvariantSL2}\eeq
We have 
\beq  \textrm{pr}^{\lf 2 \rg}\, X\lf f\rg \, I_1 = \textrm{pr}^{\lf 2 \rg}\, Y \lf g \rg\, I_1 = 0   \eeq for arbitrary $f$ and $g$, but 
\beq  \textrm{pr}^{\lf 2 \rg}\, X\lf f\rg \, I_2 =  \frac{2 f_{xxx} \left(3 u_y^2-2 u u_{yy}\right)}{u^4}, \quad 
\textrm{pr}^{\lf 2 \rg}\, Y\lf g \rg \, I_2 =  \frac{2 g_{yyy} \left(3 u_x^2-2 u u_{xx}\right)}{u^4} . \eeq
Thus, $I_1$ is invariant under the direct product the two  Virasoro groups $ VIR\lf x \rg \otimes VIR\lf y \rg$. The PDE $I_1 = A$ , for any real constant $A$, is invariant under this group. For $A \neq 0$ we scale to $A= 1$ and obtain the equation (\ref{AlgLiouvilleEq}). For $A =0 $ we obtain an equation equivalent to the linear wave equation $z_{xy}=0$, namely
\beq u u_{xy}-u_x u_y = 0.\eeq

On the other hand $I_2$ is invariant only for $f_{x x x} = g_{y y y} = 0$, i.e. it is only invariant under  $ SL_x \lf 2 , \mathbb{R} \rg \otimes SL_y \lf 2 , \mathbb{R} \rg $. Even the equation $I_2 =0$  is only invariant on the manifold satisfying the system 
\beq 2 u u_{x x} - 3 u_x^2 = 0 , \qquad  2 u u_{y y} - 3 u_y^2 = 0 ,\label{weakInv}\eeq
i.e. on a very restricted class of solutions, namely 
\beq u = \lf  a \, x \, y + b \, x +c \, y +d \rg ^{-2}  , \eeq
for arbitrary constants $a, \dots , d$. 
%%%%%%%%%%%%%%%%%%%%%%%%%
\section{Symmetry preserving discretization of partial difference equations.}
The basic idea of the invariant discretization of a PDE  is to replace it  by a system of difference equations, formed out of invariants of the action of the symmetry group of the PDE. This difference system ($\Delta$S) describes both the original PDE and a lattice \cite{d91,Doro-book,LW2006,W2004,WCUP2011}.

To be specific, let us restrict to the case of one scalar PDE involving two independent variables ($x,y$) and one dependent one $u(x,y)$. The PDE is 
\beq F\lf x, y, u, u_x, u_y, u_{xx}, u_{xy}, u_{yy}, \cdots \rg=0 \label{3.1}\eeq
and its Lie point symmetry group $\mathcal G$ is assumed to be known, together with its symmetry algebra $\mathcal L$. The $\Delta$S describing (\ref{3.1}) will have the form
\beqa \label{3.2} E_\alpha \lf x_{m+i, n+j}, y_{m+i, n+j}, u_{m+i, n+j}   \rg =0, \\
\alpha =1, \dots, N, \quad i_{min} \leq i \leq i_{max}, \quad j_{min} \leq j \leq j_{max}. \nonumber \eeqa

On Fig.1 we depict a general lattice, a priori extending indefinitely in all directions. An orthogonal lattice (not necessarily uniform) is obtained by setting $\epsilon_{ik}=0, \, \delta_{ik}=0$. 
%%%%%%%%%%%%%%  Gray figure   %%%%%%%%%%%%%%%%%
\begin{figure}\begin{center}\includegraphics[width=0.9\textwidth]{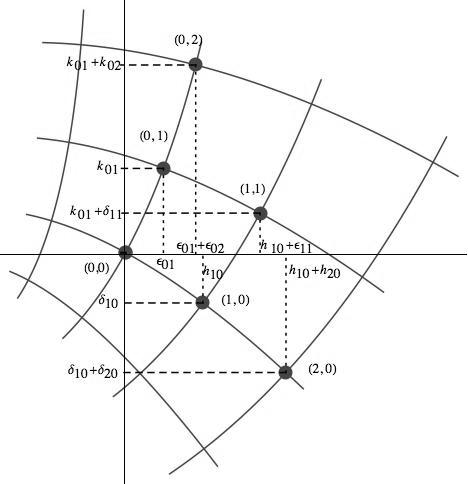}\caption{Points on a general lattice, e.g.  $x_{0,0}=x$, $x_{1,0}=x+h_{1,0}$, $x_{0,1}=x+\epsilon_{0,1}$, $x_{1,1}=x+h_{1,0}+\epsilon_{1,1}$, $x_{2,0}=x+h_{1,0}+h_{2,0}$, $x_{0,2}=x+\epsilon_{0,1}+\epsilon_{0,2}$, $y_{0,0}=y$, $y_{0,1}=y+k_{0,1}$, $y_{1,0}=y+\delta_{1,0}$, $y_{1,1}=y+k_{0,1}+\delta_{1,1}$, $y_{0,2}=y+k_{0,1}+k_{0,2}$, $y_{2,0}=y+\delta_{1,0}+\delta_{2,0}$. }\end{center}\end{figure}
%%%%%%%%%%%%%%%%%%%%%%%%%%%%%%%%%
%%%%%%%%%% Color figure %%%%%%%%%%%%%%
%\begin{figure}\begin{center}\includegraphics[width=0.9\textwidth]{fig_1-2-n}\caption{Points on a general lattice, e.g.  $x_{0,0}=x$, $x_{1,0}=x+h_{1,0}$, $x_{0,1}=x+\epsilon_{0,1}$, $x_{1,1}=x+h_{1,0}+\epsilon_{1,1}$, $x_{2,0}=x+h_{1,0}+h_{2,0}$, $x_{0,2}=x+\epsilon_{0,1}+\epsilon_{0,2}$, $y_{0,0}=y$, $y_{0,1}=y+k_{0,1}$, $y_{1,0}=y+\delta_{1,0}$, $y_{1,1}=y+k_{0,1}+\delta_{1,1}$, $y_{0,2}=y+k_{0,1}+k_{0,2}$, $y_{2,0}=y+\delta_{1,0}+\delta_{2,0}$. }\end{center}\end{figure}
%%%%%%%%%%%%%%%%%%%%%%%%%%%%%%%%
The difference system (\ref{3.2}) is written on a {\it stencil}: a finite number $N$ of adjacent points, sufficient to reproduce, in the continuous limit, all derivatives figuring in the differential equation (\ref{3.1}). For instance, for a first order PDE the minimal number of points on a stencil is three: $(m,n)$ $(m+1,n)$ $(m,n+1)$. Since the system (\ref{3.2}) is {\it autonomous}, i.e. the labels $(m,n)$ do not figure in the $\Delta$S (\ref{3.2}) explicitly, we can shift the stencil around on the lattice arbitrarily. For convenience we will choose the reference point to be $(m,n)=(0,0)$ and build the stencil around it. Thus, in (\ref{3.2}) we start with $m=n=0$ and then shift as needed.

For a first order initial value problem 
\beq F\lf x, y, u, u_x, u_y \rg=0, \quad u(x,0)=\phi(x)  \label{3.3}\eeq
it would be sufficient to choose $N=3$ in (\ref{3.2}) and give as initial data $x_{m,0}, \, y_{m,0}, \, u_{m,0}$ for all $m$. 

On the first stencil we know $x_{0,0},\,x_{1,0},\,y_{0,0},\,y_{1,0},\,u_{0,0},\,u_{1,0}$ and calculate $x_{0,1},\,y_{0,1},\,u_{0,1}$ from (\ref{3.2}). Then we shift the stencil one step in any direction and calculate further values till we fill the entire lattice.

To facilitate the calculations of the continuous limit we perform a transformation of variables on the stencil, introducing differences between coordinates and discrete partial derivatives \cite{LW2006,ltw,lstw,LR2}. The new coordinates are $\{ x_{0,0},\,y_{0,0},\,u_{0,0},\,h_{1,0},\,\epsilon_{0,1},\,k_{0,1},\,\delta_{1,0},\,u^d_x,\, u^d_y \}$, with
\beqa h_{1,0}&=&x_{1,0}-x_{0,0}, \, k_{0,1}=y_{0,1}-y_{0,0}, \, \delta_{1,0}=y_{1,0}-y_{0,0},\, \epsilon_{0,1}=x_{0,1}-x_{0,0}, \label{3.4} \\
 \label{3.5} u_x^d&=&\frac{1}{\mathcal D} [(y_{1,0}-y_{0,0}(u_{0,1}-u_{0,0})-(y_{0,1}-y_{0,0})(u_{1,0}-u_{0,0})], \\ \nonumber u_y^d&=&\frac{1}{\mathcal D} [(x_{0,1}-x_{0,0}(u_{1,0}-u_{0,0})-(x_{1,0}-x_{0,0})(u_{0,1}-u_{0,0})], \\
 \nonumber  \mathcal D&=&\epsilon_{0,1}\delta_{1,0}-h_{1,0}k_{0,1} \ne 0.\eeqa

To describe an arbitrary second order PDE we need a stencil consisting of at least six points. A possible choice is to take points $\{ (0,0), (1,0), (0,1), (1,1), (2,0), (0,2) \}$. For PDEs of the type 
\beq \label{3.6} u_{xy}=F(x, y, u, u_x, u_y), \eeq
i.e. not involving $u_{xx}$, $u_{yy}$, it might be sufficient to take four points:  $\{ (0,0), (1,0), (0,1), (1,1) \}$.

An element of the symmetry algebra $\mathcal L$ of the PDE (\ref{3.1}) will have the form
\beq \label{3.7} \hat Z = \xi(x,y,u) \partial_x + \eta(x,y,u) \partial_y + \phi(x,y,u) \partial_u 
\eeq 
where the smooth functions $\xi$, $\eta$ and $\phi$  are known (obtained by a standard algorithm for PDEs \cite{olver}).

In order to obtain an {\it invariant} $\Delta$S (\ref{3.2}) we must construct it out of difference invariants of the group $\mathcal G$, the Lie point symmetry group of the PDE (\ref{3.1}). To calculate these invariants we consider the action of the vector field $\hat Z$ at some reference point $\{ x_{0, 0}, \, y_{0, 0},\, u_{0, 0} \}$ and prolong  it to all points figuring on a chosen stencil. This amounts to a prolongation to the discrete jet space:
\beq \label{3.8}
\mbox{pr}\hat Z = \sum_{i, j}(\xi_{i, j} \partial_{x_{i, j}}+\eta_{i, j} \partial_{y_{i, j}}+\phi_{i, j} \partial_{u_{i, j}}).
\eeq
As in the continuous case,  we can use both {\it strong} and {\it weak} invariants. The strong and weak invariants satisfy
\beqa \label{3.9a} &\mbox{pr} \hat Z I_s=0,
\\
 \label{3.9b} &\mbox{pr} \hat Z I_w\left |_{I_w=0}=0,\right.
\eeqa
respectively. To determine both types of invariants we choose a basis $ \lgr \hat Z_1,\;  \cdots, \; \hat Z_A\rgr $ ($A=\mbox{dim}\mathcal L$)  for the Lie algebra $\mathcal L$  and solve the set of equations 
\beq \label{3.10} \mbox{pr} \hat Z_a I(x_{i, j},y_{i, j},u_{i, j})=0, \quad a=1,\cdots, A.
\eeq
For strong invariants the rank $r$ of the matrix of coefficients in (\ref{3.10}) is maximal and the same for all points ($m+j,n+k$). Invariants exist if we have $r=A<N$. Weak invariants are only invariant on some manifold in the space of points, obtained by requiring that the rank of coefficients in (\ref{3.10}) be less than maximal. Thus, there may be more weak invariants than strong ones (strong invariants satisfy both (\ref{3.9a}) and (\ref{3.9b})). The number of strong invariants is n = N-A.

%%%%%%%%%%%%%%%%%%%
\section{Invariant discretization  of the algebraic Liouville equation on a four-point stencil}
%%%%%%%%%%%%%%%%%%%%%%%%%%%
 We choose the four points $ \textswab{s}_4^{\,0} \equiv \lgr \lf 0, 0\rg, \lf 0, 1\rg, \lf 1, 0\rg, \lf 1, 1\rg \rgr$ on Fig.1 and can  translate them to any  stencil  $  \textswab{s}_4^{m,n} =  \{ (m,n)\,(m+1,n)\,(m,n+1)\,(m+1,n+1)\}  $ on the ($x,y$) plane.
 The vector fields  (\ref{genAlg}) of the symmetry algebra $\mathcal L$ can be discretized and prolonged to all points  of the stencil:
\beqa X^D\lf f \rg=\textrm{pr}\, X\lf f \rg &=& \sum_{\lf m, n\rg \in \textswab{s}_4^{m,n}} \lq f\lf x_{m, n}\rg \p_{x_{m n}} - f' \lf x_{m, n} \rg \, u_{m n} \, \p_{u_{m n}}   \rq ,\nn \\
Y^D\lf g \rg = \textrm{pr}\, Y\lf g \rg &=& \sum_{\lf m, n\rg \in \textswab{s}_4^{m,n}} \lq g\lf y_{m, n}\rg \p_{y_{m n}} - g\dot{{}}\lf y_{m, n} \rg \, u_{m n} \, \p_{u_{m n}}   \rq. \label{ProlDiscrVectField}\eeqa The prime and the dot denote (continuous) derivatives with respect to $x$ and $y$, respectively.

Let us first restrict to the maximal finite-dimensional subalgebra  $ sl_x \lf 2 , \mathbb{R} \rg \bigoplus sl_y \lf 2 , \mathbb{R} \rg $. The corresponding group acts transitively on the space of the continuous variables $\lf x, y, u\rg \in \mathbb{R}^3$, and   sweeps out an orbit of codimension 6 on the 12-dimensional direct product  $ \mathbb{R}^3 \bigotimes \textswab{s}_4 $. Hence we obtain 6 functionally independent invariants.  A simple basis for these invariants is given by 
\beqa \nonumber
&\xi_1=  \frac{\left(x_{0,1}-x_{0,0}\right)
   \left(x_{1,1}-x_{1,0}\right)}{\left(x_{0,0}-x_{1,0}\right)
   \left(x_{0,1}-x_{1,1}\right)} =\frac{\epsilon_{0,1} \epsilon_{1,1}}{h_{1,0}(h_{1,0}+\epsilon_{1,1}-\epsilon_{0,1})},  \\
&\eta_1=  \frac{\left(y_{0,0}-y_{1,0}\right)
   \left(y_{0,1}-y_{1,1}\right)}{\left(y_{0,1}-y_{0,0}\right)
   \left(y_{1,1}-y_{1,0}\right)} =\frac{\delta_{1,0} \delta_{1,1}}{k_{0,1}(k_{0,1}+\delta_{1,1}-\delta_{1,0})}\label{4pxyInv} \\ \nn\\
&\begin{array}{c} 
H_1= u_{0,0} u_{0,1} \epsilon_{0,1}^2 k_{0,1}^2
   \\ 
H_2 =  u_{1,0} u_{1,1} \epsilon_{1,1}^2 (k_{0,1} +\delta_{1,1}-\delta_{1,0})^2 \\ \\
H_3 =  \frac{u_{1,0} (h_{1,0}-\epsilon_{0,1})^2
   (k_{0,1}-\delta_{1,0})^2}{u_{0,0}\,
   \epsilon_{0,1}^2 \, k_{0,1}^2}
   \\ \\
H_4 =   \frac{u_{1,1} \epsilon_{1,1}^2
   (k_{0,1} + \delta_{1,1}-\delta_{1,0})^2}{u_{0,0}\,
   h_{1,0}^2 \, \delta_{1,0}^2}
\end{array} \label{4puuInv} \eeqa
The quantities $h_{1,0}$, $k_{0,1}$, $\epsilon_{0,1}$, $\epsilon_{1,1}$, $\delta_{1,0}$ and $\delta_{1,1}$ are defined on Fig. 1. The invariants $\xi_1$ and $\eta_1$ can be conveniently used to define an invariant lattice, e.g. by putting $\xi_1=A$, 
$\eta_1=B$, where $A$ and $B$ are constants. We choose the simplest possibility, namely 
\beq \label{4.4}
\xi_1=0,\qquad \eta_1=0.
\eeq
This implies that e.g. $x_{0,1}-x_{0,0}=\epsilon_{0,1}=0$ and also  as a consequence  $x_{1,1}-x_{1,0}=\epsilon_{1,1}=0$. Similarly $\delta_{1,0}=\delta_{1,1}=0$. Thus we have
\beq \label{4.5}
x_{m,n}=x_m, \qquad y_{m,n}=y_n,
\eeq
i.e. $x_{m,n}$ depends only on the first index, $y_{m,n}$ only on the second one.  We thus obtain an orthogonal lattice (in an invariant manner). The quantities $\xi_1$ and $\eta_1$ are only invariant under $SL_x(2) \otimes SL_y(2)$, however we have
\beqa \label{4.6}
&&\hat X^D(x^3) \xi_1
= (x_{1,1}-x_{0,0})(x_{1,0}-x_{0,1}) \xi_1 \left |_{\xi_1=0} =0 \right. \\ \nonumber
&&\hat X^D(x^3) \eta_1=0.
\eeqa
It follows from the commutation relations (\ref{commrel}) that a quantity annihilated by $\hat X^D(x^3)$ is also annihilated by $\hat X^D(x^n)$ for any $n$. Thus the lattice condition (\ref{4.4}) is invariant under $VIR (x) \otimes VIR (y)$. On the other hand the equations $\xi_1=A$, 
$\eta_1=B$, where $A$ and $B$ are nonzero constants are not Virasoro invariant. We conclude that
an orthogonal lattice is obligatory if we define it in terms of $\xi_1$ and $\eta_1$  alone.
Conditions (\ref{4.4}) and (\ref{4.5}) are compatible with choosing a uniform orthogonal lattice 
\beq \label{4.7}
x_m = a m + x_0, \qquad y_n = b n + y_0,
\eeq 
where $a > 0$, $b > 0$, $x_0$, $y_0$ are constants, but this choice is not obligatory.

The invariants $H_1, \cdots, H_4$ of (\ref{4puuInv}) are not suitable on the lattice (\ref{4.4}) since they all vanish or become infinite on the lattice. Before specifying the lattice we must choose new invariants (functions of those in (\ref{4pxyInv}) and (\ref{4puuInv})) which remain finite and nonzero for $\epsilon_{i,j}=\delta_{i,j}=0$. Only two such $SL_x(2) \otimes SL_y(2)$ invariants exist, namely:
\beqa \label{4.8}
J_1 &=& H_1 H_3 = u_{0,1} u_{1,0} h_{1,0}^2 k_{0,1}^2, \\ \label{4.9}
J_2 &=& \frac{1}{\xi_1^2}\frac{H_2}{H_3} = u_{0,0} u_{1,1} h_{1,0}^2 k_{0,1}^2.
\eeqa
Neither of them is strongly  invariant under the Virasoro group, since we have
\beq \label{4.10}
\hat X^D(x^3) J_1 = - h_{1,0}^2 J_1, \qquad \hat X^D(x^3) J_2 = - h_{1,0}^2 J_2.
\eeq
The equation $J_2-J_1=0$ is Virasoro invariant (on its solution set) and this equation is a discretization of $u u_{xy}-u_x u_y =0$ (equivalent to the wave equation $z_{xy}=0$).

Putting $u_{0,0}=u(x,y)$, $u_{1,0}=u(x+h_{1,0},y)$, $u_{0,1}=u(x,y+k_{0,1})$ and $u_{1,1}=u(x+h_{1,0},y+k_{0,1})$,  expanding in a Taylor series and keeping only the lowest order terms, we find
\beq \label{4.11}
J_2-J_1=h_{1,0}^3 k_{0,1}^3 (u u_{xy}-u_x u_y ).
\eeq
The Liouville equation is approximated by the difference scheme
\beqa \label{4.12}
J_2 - J_1 = a |\,J_1\, |^{3/2} + b J_1 |\,J_2\, |^{1/2} + c |\,J_1\, |^{1/2} J_2 + d |\,J_2\, |^{3/2},
\\ \nonumber  \xi_1 =0, \quad \eta_1 =0, \qquad a+b+c+d=1. 
\eeqa
Indeed the Taylor expansion yields
\beqa \label{4.13}
&& J_2 - J_1 -\left [ a J_1^{3/2} + b J_1 I_2^{1/2} + c J_1^{1/2} J_2 + d J_2^{3/2} \right ] = \\
\nonumber && = h_{1,0}^3 k_{0,1}^3 \left [ u u_{xy}-u_x u_y-u^3 \right ] + h_{1,0}^4 k_{0,1}^3 \left [ \frac 1 2  u_y u_{xx}(u - 1) - \frac 3 2 u^2 u_x \right ] + \\ \nonumber && + h_{1,0}^3 k_{0,1}^4 \left [ \frac 1 2  u_x u_{yy}(u - 1) - \frac 3 2 u^2 u_y \right ] +\mathcal O(h_{1,0}^4 k_{0,1}^4),
\eeqa
where the constant $a, \; b, \; c, \; d$ appear in the $O(h_{1,0}^4 k_{0,1}^4)$ terms.
The $\Delta$S (\ref{4.12}) is $SL_x(2) \otimes SL_y(2)$ invariant, not however Virasoro invariant. The scheme is suited for solving a boundary value problem. Give ($x,y$) in the points ($m,0$), ($0,n$) then start from ($0,0$), ($1,0$), ($0,1$) and calculate ($x_{1,1}, y_{1,1}, u_{1,1}$). Then move the stencil up or to the right and cover the entire first quadrant in the computational space ($m,n$).

%%%%%%%%%%%%%%%%%%%%%%%%%%%%%%%
  \section{Numerical results and analysis }
  %%%%%%%%%%%%%%%%%%%%%%%%%%%%%%%
In order to to test the efficiency of the numerical algorithms based on the invariant difference scheme (\ref{4.12}), we will solve a set of boundary value problems for the Liouville equation on a uniform lattice $h_{m,n} = h, \; k_{m,n} = k$. Then, we will  compare the results with the analytic solutions and with the corresponding ones obtained by the  standard finite difference approximation 
\beq u_{1,1} u_{0,0}-u_{0,1} u_{1,0} =  h k \; u_{0,0}^3. \label{stand-discr} \eeq
Both the equations (\ref{4.12}) and (\ref{stand-discr}) relate the values at the corner of a rectangle of meshes of length   $h$ and $k$, respectively. Then a natural class of boundary value problems  consists in giving  the value of $u$  on two  sets of points of the form $ \lf m, 0 \rg $ and $ \lf 0, n\rg  $  for $m, n \in \mathbb{N }$  in the computational basis.  
%%%%%%%%%%%%%%  Gray Figure   %%%%%%%%%%%%%%%%
%\begin{figure}\begin{center}\includegraphics[width=0.7\textwidth]{NewInvariants-gr}\caption{In the 4 point scheme, adopted both in the standard discretization of the Liouville equation (\ref{stand-discr}) and in the invariant  discretization (\ref{4.12}), the value of $u$  at the right top point in each rectangle is obtained  using the values in  the three other vertices. In the considered   boundary value problem, the values of $u$ are given in the points (m,0) and (0,n). Then, starting from the rectangle  at the left bottom corner, denoted by $0$, one gets the value $u_{11}$ from the data connected by the dotted diagonal.  This can be used to evaluate the  right top point of the rectangle denoted by $1$ together with the  data in $\lf 1,0\rg$ and $\lf 2, 0\rg$. Proceed further, till the first row of rectangles is completed, then repeat the same procedure for the second row, involving also the data in $\lf 0, 2\rg$.  In the figure are indicated the pair of points involved in the computation of the invariants in each rectangle.  }\end{center}\end{figure}
%%%%%%%%%%%%%%%%%%%%%%%%%%%%%%%%%%%%%%
%%%%%%%%%%%%%%  Color  Figure   %%%%%%%%%%%%%%%%
\begin{figure}\begin{center}\includegraphics[width=0.7\textwidth]{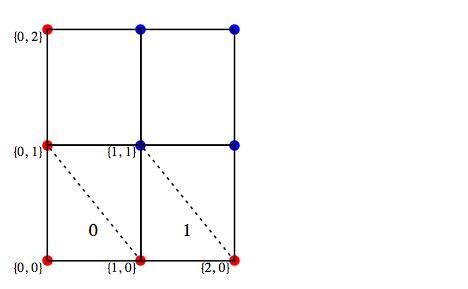}\caption{In the 4 point scheme, adopted both in the standard discretization of the Liouville equation (\ref{stand-discr}) and in the invariant  discretization (\ref{4.12}), the value of $u$  at the right top point in each rectangle is obtained  using the values in  the three other vertices. In the considered   boundary value problem, the values of $u$ are given in the points (m,0) and (0,n). Then, starting from the rectangle  at the left bottom corner, denoted by $0$, one gets the value $u_{11}$ from the data connected by the dotted diagonal.  This can be used to evaluate the  right top point of the rectangle denoted by $1$ together with the  data in $\lf 1,0\rg$ and $\lf 2, 0\rg$. Proceed further, till the first row of rectangles is completed, then repeat the same procedure for the second row, involving also the data in $\lf 0, 2\rg$.  In the figure are indicated the pair of points involved in the computation of the invariants in each rectangle.  }\end{center}\end{figure}
%%%%%%%%%%%%%%%%%%%%%%%%%%%%%%%%%%%%%%%%
\bigskip
%%%%%%%%%%%%%%  Gray Figure   %%%%%%%%%%%%%%%%
%\begin{figure}[h]\begin{center}$\begin{array}{cc}a) \includegraphics[width=2.5in]{QLoren-numerics-gr} & b) \includegraphics[width=2.5in]{QLoren-rel-gr} \\ c) \includegraphics[width=2.5in]{QLoren-numerics-standard-gr} & d) \includegraphics[width=2.5in]{QLoren-rel-standard-gr}\end{array}$\end{center}\caption{The solution $s_1$ with the choice of parameters $\alpha=6,\; \beta = 1,\;\gamma = 1,\; \delta = 1$  is numerically computed giving a boundary value problem on a lattice with corner point $\lf x_{0 0}, y_{0 0}\rg  = \lf -2.5, -2.5\rg $ and steps of equal length $ h = k= 0.02 $ for a lattice of $260 \times 260 $ points.  Numerical results using the invariant formula (\ref{4psymminvfor})   are shown in $a)$, and the relative error with respect to the analytic solution in $b)$. Analogously, numerical results obtained by the standard formula (\ref{stand-discr}) are reported in $c)$ and the corresponding relative error in $d)$. Despite  the generic similarities of the two results,  the difference of two orders of magnitude in the relative errors is remarkable.  }\end{figure}
%%%%%%%%%%%%%%%%%%%%%%%%%%%%%%%%%%%%%%%%%%%%%%
%%%%%%%%%%%%%%  Color Figure   %%%%%%%%%%%%%%%%
\begin{figure}[h]\begin{center}$\begin{array}{cc}a) \includegraphics[width=2.5in]{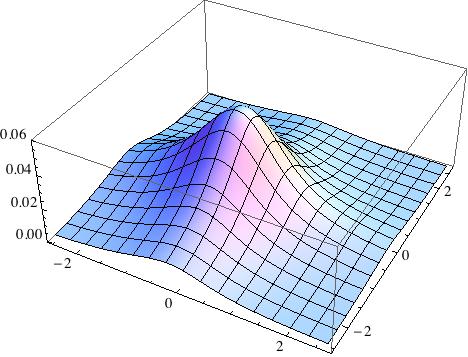} & b) \includegraphics[width=2.5in]{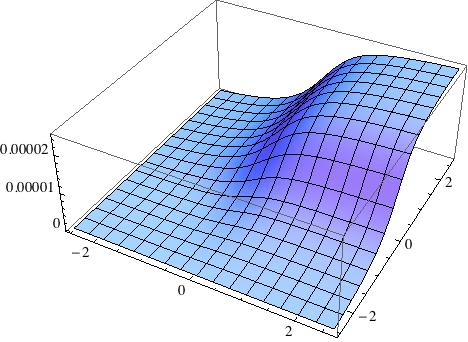} \\ c) \includegraphics[width=2.5in]{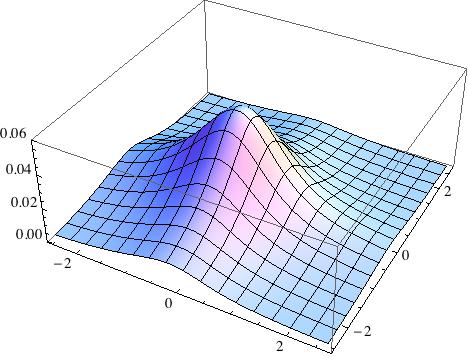} & d) \includegraphics[width=2.5in]{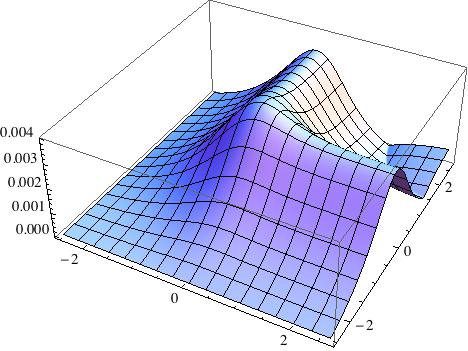}\end{array}$\end{center}\caption{The solution $s_1$ with the choice of parameters $\alpha=6,\; \beta = 1,\;\gamma = 1,\; \delta = 1$  is numerically computed giving a boundary value problem on a lattice with corner point $\lf x_{0 0}, y_{0 0}\rg  = \lf -2.5, -2.5\rg $ and steps of equal length $ h = k= 0.02 $ for a lattice of $260 \times 260 $ points.  Numerical results using the invariant formula (\ref{4psymminvfor})   are shown in $a)$, and the relative error with respect to the analytic solution in $b)$. Analogously, numerical results obtained by the standard formula (\ref{stand-discr}) are reported in $c)$ and the corresponding relative error in $d)$. Despite  the generic similarities of the two results,  the difference of two orders of magnitude in the relative errors is remarkable.  }\end{figure}
%%%%%%%%%%%%%%%%%%%%%%%%%%%%%%%%%%%%%%%%%%%%%%
Thus, one can proceed in calculating the fourth value of u from three given values on each rectangle, as depicted on Fig.2, starting from the left bottom one at the corner. The problem with the formula (\ref{4.12}) is that it involves algebraic functions. However, a possibility is to make a special choice for the parameters, namely set $b = d = 0$, which leads to a linear equation for $u_{11}$ and hence to an explicit scheme. More precisely, we have a 1-parameter family of  recursion formulae 
\beq u_{1,1} = \frac{u_{0,1} u_{1,0} \left(a h k \sqrt{u_{0,1}
   u_{1,0}}+1\right)}{u_{0,0} \left((a-1) h k \sqrt{u_{0,1}
   u_{1,0}}+1\right)} \qquad \lf a \neq 0,1\rg,  \label{4psymminvfor}\eeq
   for arbitrary real $a$ ( with $c = 1 - a$). Furthermore, to simplify calculations we require that  the unknown function is  strictly positive. In the actual calculations we used the symmetric case $a = c = \half$.
   
   We used different  exact solutions of the Liouville equations, among them for instance
   \beqa 
   s_1 &=&  \frac{2  \beta \gamma  \delta }{\left(\beta^2 x^2+1\right) \left(\delta^2
   y^2+1\right) \left( \tan ^{-1}(\beta x)+\gamma \tan
   ^{-1}(d y)+\alpha\right)^2},  \\  s_2 &=& \frac{2 A s^2 e^{s (x+y)}}{\left(A e^{s y}+e^{s x}\right)^2},  \\
    s_3 &=&  \frac{8 \left(1-4 \left(x+\frac{1}{2}\right)\right) (1-4 y)
   \exp \left(-4 \left(x+\frac{1}{2}\right)^2+2
   \left(x+\frac{1}{2}\right)-4 y^2+2 y\right)}{\left(e^{2
   \left(x+\frac{1}{2}\right)-4
   \left(x+\frac{1}{2}\right)^2}+e^{2 y-4 y^2}+1\right)^2} ,\label{f3}
    \eeqa
for certain values of the constants $A, s , \alpha, \beta, \gamma, \delta $.  Once  the values for the lattice constants $h$ and $k$ and the corner point $\lf 0, 0 \rg$ are fixed the values of the analytic solution on the points of the boundary are computed and used as initial data for the numerical calculations. For some of the  functions defined above, both the invariant formula (\ref{4psymminvfor}) and the standard formula (\ref{stand-discr}) are used to compute the  solutions and compare them with the known analytically computed values at the lattice points. As an illustrative example, in Figure 3, we report the calculations made for the solution $s_1$.

Supplementary material of the same kind is provided in Fig.4 and Fig. 5 for  $s_2$ and  $s_3$. In all cases the agreement with the exact formulas is much better  for the invariant schemes than for the standard ones.

In order to provide a rough evaluation of how correctly the numerical calculations reproduce the analytical solutions below we give a table,  where the distances,  as mean square  averages (or the normalized  $L^2_{\mathbb{R}^2}$ metric),  between  the numerical solutions computed by the invariant scheme and the standard method, respectively,  w.r.t. the analytic ones are compared:
\begin{center} \begin{tabular}{ l | c r }
   & $\chi_{Inv}$  & $\chi_{stand} \qquad$    \\ \hline  
  $s_1$ & $6.4 \times 10^{-16}$ & $ 7.2 \times 10^{-5}$ \\
  $s_2$ & $1.6 \times 10^{-7}$ & $7.0 \times 10^{-1}$  \\
   $s_3$ &  $1.7 \times 10^{-2}$  & $6.0 \times 10^{-1}$  \\
\end{tabular}  \end{center}
%%%%%%%%%%%%%%  Gray Figure   %%%%%%%%%%%%%%%%
%\begin{figure}[h]\begin{center}$\begin{array}{cc} a) \includegraphics[width=2.5in]{qGaussian-numerics-gr} & b) \includegraphics[width=2.5in]{qGaussian-rel-gr} \\  c) \includegraphics[width=2.5in]{qGaussian-numerics-standard-gr} & d) \includegraphics[width=2.5in]{qGaussian-rel-standard-gr}\end{array}$\end{center}\caption{The same analysis as above  for the solution $s_2$  on a lattice with corner point $\lf x_{0 0}, y_{0 0}\rg  = \lf -1.5, -1.0\rg $ and steps of equal length $ h = k= 0.02$ for a lattice of $60 \times 60$ points.  }\end{figure}
%%%%%%%%%%%%%%%%%%%%%%%%%%%%%%%%%%%%%%%%%%%%%%%%%%%
%%%%%%%%%%%%%%  Color Figure   %%%%%%%%%%%%%%%%
\begin{figure}[h]\begin{center}$\begin{array}{cc} a) \includegraphics[width=2.5in]{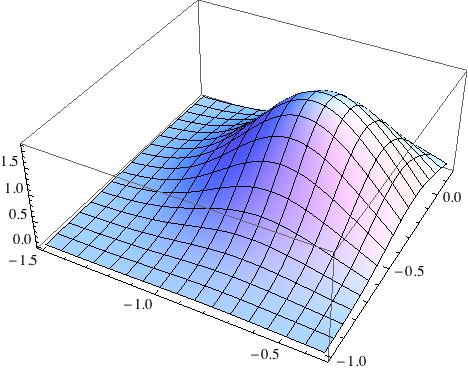} & b) \includegraphics[width=2.5in]{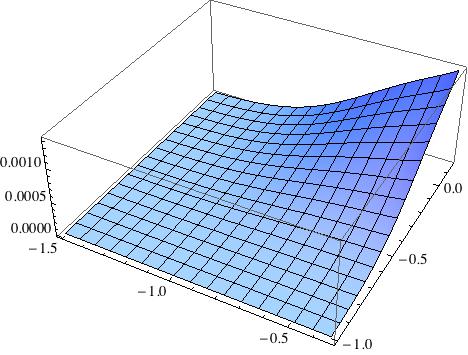} \\  c) \includegraphics[width=2.5in]{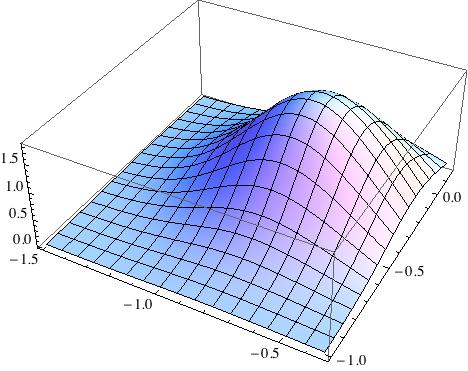} & d) \includegraphics[width=2.5in]{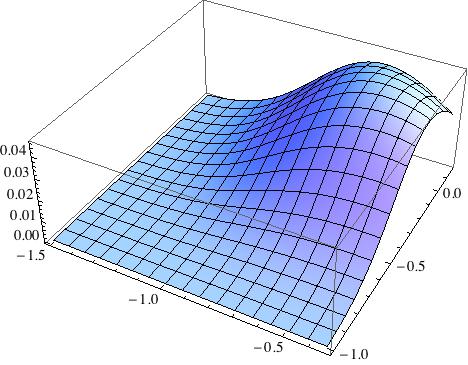}\end{array}$\end{center}\caption{The same analysis as above  for the solution $s_2$  on a lattice with corner point $\lf x_{0 0}, y_{0 0}\rg  = \lf -1.5, -1.0\rg $ and steps of equal length $ h = k= 0.02$ for a lattice of $60 \times 60$ points.  }\end{figure}
%%%%%%%%%%%%%%%%%%%%%%%%%%%%%%%%%%%%%%%%%%%%%%%%%%%
%%%%%%%%%%%%%%  Gray Figure   %%%%%%%%%%%%%%%%
%\begin{figure}[]\begin{center}$\begin{array}{cc}a) \includegraphics[width=2.5in]{Wall-2-numerics-gr} &b) \includegraphics[width=2.5in]{Wall-2-rel-gr} \\ c) \includegraphics[width=2.5in]{Wall-2-numerics-standard-gr} &d) \includegraphics[width=2.5in]{Wall-2-rel-standard-gr}\end{array}$\end{center}\caption{The same analysis as above  for the solution $s_3$ for the choice of parameters $A = 12.8397, s = 3.86233$  on a lattice with corner point $\lf x_{0 0}, y_{0 0}\rg  = \lf -3, -1\rg $ and steps of equal length $ h = k= 0.02$ for a lattice of $180 \times 180$ points.} \end{figure}
%%%%%%%%%%%%%%%%%%%%%%%%%%%%%%%%%%%%%%%%%%%%%%%%%%%
%%%%%%%%%%%%%%  Color  Figure   %%%%%%%%%%%%%%%%
\begin{figure}[]\begin{center}$\begin{array}{cc}a) \includegraphics[width=2.5in]{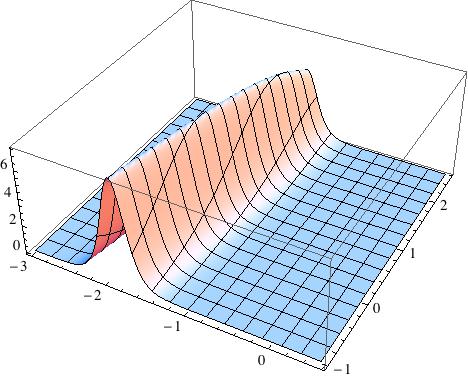} &b) \includegraphics[width=2.5in]{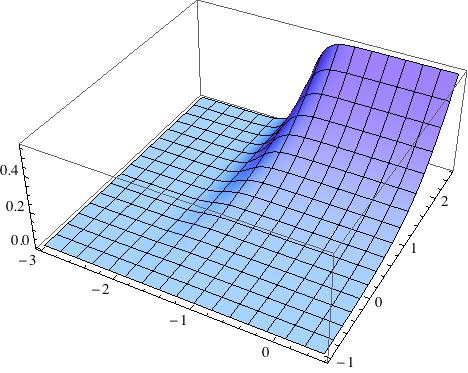} \\ c) \includegraphics[width=2.5in]{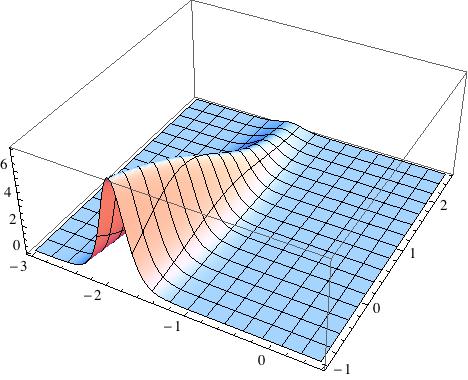} &d) \includegraphics[width=2.5in]{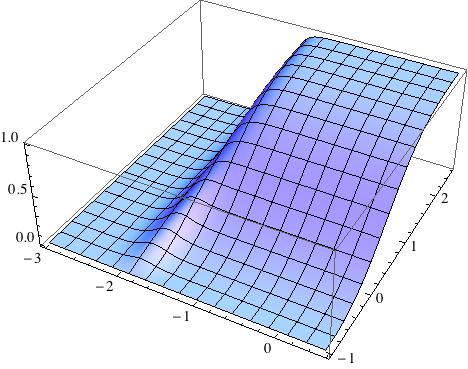}\end{array}$\end{center}\caption{The same analysis as above  for the solution $s_3$ for the choice of parameters $A = 12.8397, s = 3.86233$  on a lattice with corner point $\lf x_{0 0}, y_{0 0}\rg  = \lf -3, -1\rg $ and steps of equal length $ h = k= 0.02$ for a lattice of $180 \times 180$ points.} \end{figure}
%%%%%%%%%%%%%%%%%%%%%%%%%%%%%%%%%%%%%%%%%%%%%%%%%%%

 \newpage
\section{Symmetries of Rebelo-Valiquette Liouville discretized  equation}
%%%%%%%%%%%%%%%%%%%%%%%%%%%%%%%%%%
In  \cite{RebeloValiquette}  Rebelo and Valiquette considered a symmetry preserving discretization of the Liouville equation \eqref{AlgLiouvilleEq}  namely:
\beqa \label{Li4}
&&L_{RV}^D = u_{11}u_{00} - u_{10}v_{01} - u_{00}u_{01}u_{10}(x_{10}-x_{00})(y_{01}-y_{00}) = 0, \\ \nonumber
&&\quad x_{01}=x_{00}, \quad y_{10}=y_{00}.
\eeqa
The equation for the lattice clearly states that $x_{i j} = x_i$ and $y_{i j} = y_j$, so the lattice coincides with the one we used above. 
They constructed (\ref{Li4}) from the invariance with respect to  the pseudo--group 
\beqa \label{Li4a}
\tilde x_{i}=F(x_i), \quad \tilde y_j=G(y_j), \quad \tilde u_{i j}=\frac{u_{i j}}{\frac{F(x_{i+1})-F(x_i)}{x_{i+1}-x_{i} } \frac{G(y_{j+1})-G(y_j)}{y_{j+1}-y_{j} }}
\eeqa for arbitrary regular $F$ and $G$.

First, let us notice that the equation (\ref{Li4}) is not invariant with respect the algebra $sl_x \lf 2 , \mathbb{R}\rg \oplus  sl_y \lf 2 , \mathbb{R}\rg$ considered in the previous sections. In fact it results that
\beq  X^D\lf x^2 \rg L_{RV}^D |_{L_{RV}^D = 0} = u_{00}u_{01}u_{10}(x_{10}-x_{00})^2(y_{01}-y_{00})  \eeq
and similarly for $Y^D\lf y^2 \rg$.

Thus, let us look here for  infinitesimal symmetries  of \eqref{Li4} of the form
\beqa \label{Li4b}
\hat X = Q^{(1)}_{ij}(x_{ij},y_{ij},u_{ij})\partial_{x_{ij}}+Q^{(2)}_{ij}(x_{ij},y_{ij},u_{ij})\partial_{y_{ij}}+Q^{(3)}_{ij}(x_{ij},x_{i+1,j},y_{ij},y_{i,j+1},u_{ij})\partial_{u_{ij}}.
\eeqa
The determining equations are:
\beqa \label{Li4c1}
&&Q^{(1)}_{01}=Q^{(1)}_{00}, \\ \label{Li4c2}
&&Q^{(2)}_{10}=Q^{(2)}_{00}, \\ \label{Li4c3}
&&Q^{(3)}_{11} u_{00}+u_{11} Q^{(3)}_{00}-u_{10}Q^{(3)}_{01}-u_{01}Q^{(3)}_{10}= Q^{(3)}_{00}u_{01}u_{10}(x_{10}-x_{00})(y_{01}-y_{00})+ \\
\nonumber &&+Q^{(3)}_{10}u_{01}u_{00}(x_{10}-x_{00})(y_{01}-y_{00})+Q^{(3)}_{01}u_{00}u_{10}(x_{10}-x_{00})(y_{01}-y_{00})+\\ \nonumber &&+u_{00}u_{01}u_{10}(Q^{(1)}_{10}-Q^{(1)}_{00})(y_{01}-y_{00})+u_{00}u_{01}u_{10}(x_{10}-x_{00})(Q^{(2)}_{01}-Q^{(2)}_{00}).
\eeqa
We put  $x_{01}=x_{00}$, $y_{10}=y_{00}$ and $u_{11}=u_{01}u_{10} \left [ \frac{1}{u_{00}}+(x_{10}-x_{00})(y_{01}-y_{00}) \right ]$ so that $x_{00}$, $y_{00}$, $y_{01}$, $x_{10}$, $u_{00}$, $u_{01}$ and $u_{10}$ are independent variables in the determining equations. From \eqref{Li4c1} we deduce that $Q^{(1)}_{ij}=f (x_i)$ and from  \eqref{Li4c2} $Q^{(2)}_{ij}=g (y_j)$ where $f$ and $g$ are arbitrary functions of their arguments.
Dividing   \eqref{Li4c3} by $u_{00}$ and applying the operator $A=u_{10} \partial_{u_{10}}-u_{01} \partial_{u_{01}}$ (we have $A \phi(u_{11})=0$ for any function $\phi$ ) and we get
\beqa \label{Li4d}
\frac{Q^{(3)}_{01}}{u_{01}}-\frac{\partial Q^{(3)}_{01}}{\partial u_{01}} = \frac{Q^{(3)}_{10}}{u_{10}}-\frac{\partial Q^{(3)}_{10}}{\partial u_{10}},
\eeqa
i.e. the quantity $\frac{Q^{(3)}_{ij}}{u_{ij}}-\frac{\partial Q^{(3)}_{ij}}{\partial u_{ij}}=h(i+j) $. So 
\beqa \label{Li4e}
 Q^{(3)}_{ij}=u_{ij} \left [ h(i+j) \log_e (u_{ij}) + A_{ij} (x_{ij},x_{i+1,j},y_{ij},y_{i,j+1})\right ].
 \eeqa
 Introducing this result into \eqref{Li4c3} and taking into account that $\log_e (u_{11})=\log_e (u_{10})+\log_e (u_{01})+ \log_e \left [ \frac{1}{u_{00}}+(x_{10}-x_{00})(y_{01}-y_{00}) \right ]$ we find from the coefficient of  $  \log_e \left [ \frac{1}{u_{00}}+(x_{10}-x_{00})(y_{01}-y_{00}) \right ]$ that $h(i+j)=0$. Thus $Q^{(3)}_{ij}=u_{ij}  A_{ij} (x_{ij},x_{i+1,j},y_{ij},y_{i,j+1})$.
 Introducing this last result into \eqref{Li4c3} we find two equations for $A_{ij} (x_{ij},x_{i+1,j},y_{ij},y_{i,j+1})$
 \beqa \label{Li4f}
&& A_{00}+A_{11}-A_{01}-A_{10}=0, \\ \label{Li4g} && A_{ij}=-\frac{f(x_{j+1})-f(x_j)}{x_{j+1}-x_{j} }-\frac{g(y_{i+1})-g(y_i)}{y_{i+1}-y_{i} }.
 \eeqa
 Eq. \eqref{Li4f} is identically satisfied by the result obtained in \eqref{Li4g} and as a consequence the symmetry algebra of the Liouville equation presented by Rebelo and Valiquette is indeed the sum of two Virasoro algebras determined by the two functions $f$ and $g$:
 \beq \hat X\lf f, g\rg  = f(x_i) \partial_{x_{i}} + g(y_j) \partial_{y_{j}} - \lq \frac{f(x_{j+1})-f(x_j)}{x_{j+1}-x_{j} } +\frac{g(y_{i+1})-g(y_i)}{y_{i+1}-y_{i} }\rq \partial_{u_{ij}}. \label{6.12}
 \eeq 
The main difference between  these generators and those in (\ref{ProlDiscrVectField}) is that in (4.1) the coefficients  of  $\partial_{u_{ij}}$  are locally dependent  on the space  points, while   two points  are involved in  (\ref{6.12}). Thus, the expression (\ref{Li4b}) has to be understood as a summation over all points of the lattice. On the contrary    (\ref{ProlDiscrVectField}) contains only finite sums over the stencil points. Thus the Rebelo--Valiquette discretization of the Liouville equation is invariant under $VIR (x) \otimes VIR (y)$, but these are generalized symmetries rather than point ones. These are actually very special generalized symmetries: The Lie algebra (\ref{6.12}) can be integrated to the finite transformations (\ref{Li4a}). These finite transformations were actually the starting point in the Rebelo-Valiquette approach.

%%%%%%%%%%%%%%%%%%%
\section{Lie point symmetries of a linearizable Liouville equation.}
%%%%%%%%%%%%%%%%%%%%%%
Adler and Startsev \cite{as1999} have presented a discretization of the algebraic Liouville equation (\ref{AlgLiouvilleEq}) on a four-point lattice, namely 
\beq                                                         \label{Li3}
    u_{i+1,j+1}(1+\frac{1}{u_{i+1,j}})(1+\frac{1}{u_{i,j+1}})u_{i,j} = 1.
\eeq
This equation is linearizable by the substitution 
\beq \label{Li3a}
    u_{i,j}= -{(v_{i+1,j}-v_{i,j})(v_{i,j+1}-v_{i,j})\over v_{i+1,j}v_{i,j+1}},
\eeq
where $v_{i,j}$ satisfies the linear equation
\beq \label{Li3b}
v_{i+1,j+1}-v_{i+1,j}-v_{i,j+1}+v_{i,j}=0.
\eeq
Hence the general solution of (\ref{Li3}) is 
\beq \label{Li3c}
    u_{i,j}= -{(c_{i+1}-c_i)(k_{j+1}-k_j)\over(c_{i+1}+k_j)(c_i+k_{j+1})},
\eeq
where $c_i,k_j$ are arbitrary functions of one index each. We restrict (\ref{Li3}) to the stencil with $i=j=1$, i.e. 
\beq \label{7.5}
E=u_{11}(u_{10}+1)(u_{01}+1)u_{00} - u_{10}u_{01}=0,
\eeq
and calculate the Lie point symmetries of this equation. The equation is autonomous, the lattice is fixed (orthogonal and uniform).  Hence the symmetry algebra is generated by vector fields of the form 
\beq \label{sLi3}
\hat X_e=Q_{i j}(u_{ij}) \partial_{u_{ij}},
\eeq
satisfying 
\beq \label{7.7}
\hat X E |_{E=0}=0.
\eeq
We obtain 
\beqa \label{sLi3a}
&&Q_{11}(u_{01}+1)(u_{10}+1)u_{00} + Q_{10} u_{11}(u_{01}+1)u_{00} + Q_{01}u_{11}(u_{10}+1) u_{00} + \\ \nonumber
&&\qquad \qquad + Q_{00} u_{11}(u_{01}+1)(u_{10}+1) = Q_{10}u_{01} + Q_{01} u_{10}.
\eeqa
We eliminate $u_{11}$ from (\ref{sLi3a}) using (\ref{7.5}), then differentiate with respect to $u_{00}$ and obtain
\beq \label{sLi3c}
\frac{Q_{11}}{u_{11}}- \frac{d Q_{11}}{d u_{11}} = \frac{Q_{00}}{u_{00}}- \frac{d Q_{00}}{d u_{00}}. 
\eeq
The general solution of (\ref{sLi3c}) is 
\beq \label{7.10}
Q_{ij}=u_{ij} \left [ g_{ij} + f(i-j) \log_e(u_{ij}) \right ]
\eeq
where $g_{ij}$ and  $f(i-j)$ are functions of $i$ and $j$. Substituting (\ref{7.10}) into (\ref{sLi3a}) we find $g(i,j)=f(i-j)=0$.

It follows that the linearizable discrete Liouville equation has no continuous point symmetries at all!

Two comments are in order.
\begin{enumerate}
\item The equation (\ref{Li3}) is linearizable and hence must have generalized symmetries.
\item Ref.\cite{as1999} also contains a linearizable differential - difference Liouville equation:
\beq  \label{Li2}
    \dot u_{i+1}u_i-u_{i+1}\dot u_{i} = u_{i+1}u_i(u_{i+1}+u_i).
\eeq
where the dot denotes the derivative of $u_n(x)$ with respect to the continuous variable $x$. It can be shown using the formalism presented in \cite{LWY2002} that (\ref{Li2}) does have an infinite dimensional Lie point symmetry algebra, isomorphic to the Virasoro algebra. The algebra is realized by evolutionary vector fields of the form 
\beq \label{sLi2}
\hat X_e=Q_i(x,u_i, \dot u_i) \partial_{u_i}, \quad 
Q_i=f(x) \dot u_i + \dot f(x) u_i.
\eeq
This corresponds to the standard factor fields
\beq \label{7.13}
\hat X = f(x) \partial_x - \dot f(x) u \partial_u.
\eeq
\end{enumerate}

%%%%%%%%%%%%%%%%%%%
\section{Conclusions}\label{Conclusions}
%%%%%%%%%%%%%%%%%%%%%%
We have shown that at least on a four--point lattice it is not possible to discretize the Liouville equation (\ref{AlgLiouvilleEq}) (nor (\ref{LiouvilleEq})) while preserving $VIR (x) \otimes VIR (y)$ as the Lie point symmetry group. That is also impossible on a six--point lattice. On the other hand, Rebelo and Valiquette \cite{RebeloValiquette} have introduced a special type of generalized symmetries that leave their discretization of the algebraic Liouville equation invariant. In the continuous case these symmetries reduce to point ones. In the discrete case they are special in that the vector fields can be integrated to group transformations acting on the equation and on the lattice. This is somewhat similar to the case of the symmetries of the Toda hierarchy \cite{hlrw} where some generalized symmetries {\it contract} to point ones in the continuous limit. 

From the point of view of numerical methods it remains to explore which discretization provides better results. A discretization preserving the maximal finite subgroup of an infinite dimensional point symmetry group, or one that transforms  point symmetries into generalized ones. 

As stated in the Introduction, the main purpose of this article is to investigate how continuous physical theories can be discretized while preserving their continuous Lie point symmetries. For the Liouville equation we have shown that in a complete discretization it is possible to preserve invariance under under the maximal finite subgroup. The infinite dimensional Lie  pseudogroup does not survive as a group of point symmetries.  Rebelo and Valiquette have shown that the entire Virasoro pseudogroup does  survive in a different discretization \cite{RebeloValiquette}, but as generalized symmetries.

In Section 5 we have tested the quality of our invariant discretization as a numerical method. We have
shown that it actually performs very well. We are of course aware that what we here call "standard"
methods can be improved in many other ways. The use of point symmetries in numerical solutions of partial differential equations deserves a further detailed analysis. 

Another interesting point is that the linearizable discretization of Adler and Startsev preserves no point symmetries. It is thus important to decide which features of a continuous theory one wishes to preserve in a discretization.  In this case linearizability is incompatible with the preservation of point symmetries.

\subsection*{Acknowledgments}
DL  has been partly supported by the Italian Ministry of Education and Research, 2010 PRIN {\it Continuous and discrete nonlinear integrable evolutions: from water waves to symplectic maps}. 
\noindent LM has been partly supported by the Italian Ministry of Education and Research (PRIN 2011 {\it Teorie geometriche e analitiche dei sistemi Hamiltoniani in dimensioni finite e infinite }) and by INFN  (IS-CSN4 {\it Mathematical Methods of Nonlinear Physics}). 
\noindent The research of PW is partially supported  by a research grant from NSERC of Canada. 
 PW thanks the European Union Research Executive Agency for the award of a Marie Curie International Incoming Research Fellowship  making his stay at University Roma Tre possible. He  thanks the Department of Mathematics and Physics of Roma Tre for hospitality.

%%%%%%%%%%%%%%%%

%%%%%%%%%%%%%%%%%%%%%% 

\begin{thebibliography}{99}

\bibitem{as1999}Adler V E and Startsev S Ya 1999 Discrete analogues of the Liouville equation {\it Theor. Math. Phys.} {\bf 121}  1484--1495.
\bibitem{bcw06}Bourlioux A, Cyr-Gagnon C and Winternitz P 2006 Difference schemes with point symmetries and their numerical tests {\it J.Phys. A} {\bf 39} 6877--6896.
\bibitem{brw08}Bourlioux A, Rebelo R and Winternitz P 2008 Symmetry preserving discretization of SL(2,R) invariant equations. {\it J. Nonlinear Math. Phys.} {\bf 15 } 362--372.
\bibitem{d91}Dorodnitsyn V A 1991 Transformation groups in difference spaces, {\it J. Soviet Math.} {\bf 55} 1490--1517.
\bibitem{Doro-book} Dorodnitsyn V A 2011 {\it Applications of Lie Groups to Difference Equations}, CRC Press
\bibitem{dkw00}Dorodnitsyn V A, Kozlov R and Winternitz P 2000 Lie group classification of second--order ordinary difference equations, {\it J. Math. Phys.}  {\bf 41} 480--504.
\bibitem{hlw06}Hairer E, Wanner G and  Lubich C 2006 Geometric Numerical Integration.
Structure-Preserving Algorithms for Ordinary Differential Equations, Springer \& Verlag. 
\bibitem{hlrw}Hernandez-Heredero R, Levi D, Rodriguez M A and Winternitz P 2000 Lie algebra contractions and symmetries of the Toda hierarchy {\it J. Phys. A} {\bf 33} 5025--5040.
%\bibitem{CRCHand3}Handbook of Lie Group Analysis of Differential Equations. Vol.2: Applications in engineering and physical sciences, CRC Press, Boca Raton, 1996, N.H. Ibragimov editor .
\bibitem{i08}Iserles A 2008  {\it A First Course in the Numerical Analysis of Differential Equations} (2nd edition), Cambridge University Press. 
\bibitem{LOTW}Levi D, Olver P, Thomova Z and Winternitz P (editors) 2011 {\it Symmetries and Integrability of Difference Equations}, CUP, LMS Lecture Series.

\bibitem{lstw}Levi D, Scimiterna C, Thomova Z, and Winternitz P 2012 Contact trans- formations for difference equations. {\it J.Phys.A. Math.Theor.(Fast track communication)} {\bf 45} 022001, 9 pages.

\bibitem{ltw}Levi D, Thomova Z and Winternitz P 2011 Are there contact transformations for discrete equations? {\it J.Phys.A.Math.Theor.} {\bf 44} 265201, 7 pages.

%\bibitem{lps}D. Levi, L. Pilloni and P. M. Santini, Integrable three- dimensional lattices, {\it J.Phys. A} {\bf 14} (1981) 1567.
\bibitem{LW91}Levi D and Winternitz P 1991 Continuous symmetries of discrete equations, {\it Phys. Lett. A} {\bf 152} 335--338.
\bibitem{LW1996} Levi D and Winternitz P 1996 Symmetries of discrete dynamical systems. {\it J.Math. Phys.} {\bf 37} 5551--5576

\bibitem{LW2006} Levi D and Winternitz P 2006 Continuous symmetries of difference equations. {\it J. Phys. A}  {\bf 39}, no. 2, R1-R63
\bibitem{LR2} Levi D and  Rodriguez M A  2014  On the construction of partial  difference schemes II: discrete variables and invariant schemes  {\it arXiv:1407.0838}
\bibitem{LWY2002} Levi D, Winternitz P and Yamilov R I 2010 Lie point symmetries of
differential--difference equations. {\it J.Phys.A.Math.Theor. (Fast Track Communications)} {\bf 43} 292002,14 pages.

%\bibitem{lw97}D. Levi and R. Yamilov, Conditions for the existence of higher symmetries of evolutionary equations on the lattice, J. Math. Phys. {\bf 38} (1997) 6648-6674.

\bibitem{Liouville}Liouville J 1853 Sur l'equation aux differences partielles $\frac{d^2 \log \lambda/}{du\, dv} \pm \frac{\lambda}{2 a^2} =0$  {\it  J. Math. Pure Appl.} 1 Ser. {\bf{18}} 71--72. 

\bibitem{mw01}Marsden J E and West M 2001 Discrete mechanics and variational integrators {\it Acta Numerica} {\bf 10} 357-514.

\bibitem{mq06}McLachlan R I and Quispel G R W 2006 Geometric integrators for ODEs, {\it J. Phys. A} {\bf 39} 5251-5286.

\bibitem{Medolaghi}Medolaghi P  1898 Classificazione delle equazioni alle derivate parziali del secondo ordine, che ammettono un gruppo infinito di trasformazioni puntuali,  {\it Ann.  Mat. Pura  Appl.} {\bf{1}} 229--263.

%\bibitem{m}T. Miwa, On Hirota's difference equations, {\it Proc. Japan Acad. Ser. A Math. Sci.} {\bf 58} (1982), 1--49.

\bibitem{olver} Olver P J 1993 {\it Applications of Lie groups to differential equations}, Springer Verlag

\bibitem{rv13}Rebelo R and Valiquette F	2013 	Symmetry preserving numerical schemes for partial differential equations and their numerical tests, {\it J.	Difference Eq.  Appl.} {\bf 19} 737--757.

\bibitem{RebeloValiquette}  R Rebelo and F Valiquette 2014 Invariant Discretization of Partial Differential Equations Admitting Infinite-Dimensional Symmetry Groups, arXiv:1401.4380.

\bibitem{rw09}Rebelo R and Winternitz P  2009  Invariant difference schemes and their applications to SL(2,R) invariant differential equations  {\it J. Phys. A. Math.Theor.} {\bf 42} 454016, 10 pages (Special issue devoted to Symmetries and Integrability of Difference Equations).

\bibitem{W2004} Winternitz P 2004 Symmetries of discrete systems. In B. Grammaticos,
Y. Kosmann-Schwarzbach, and T. Tamizhmani, editors, {\it Discrete Integrable Systems}, volume 644 of Lecture Notes in Physics, pages 185--243.
Springer Verlag. See also ArXiv nlin.SI/0309058.

\bibitem{WCUP2011}Winternitz P 2011 Symmetry preserving discretization of differential equations and Lie point symmetries of differential-difference equations. In {\it Symmetries and Integrability of Difference Equations}, LMS Lecture Series, editors:Levi, Olver, Thomova, Winternitz. CUP 
%\bibitem{Yam2006} Yamilov R 2006 Symmetries as integrability criteria for differential difference equations {\it J. Phys. A: Math. Gen.} {\bf 39} R541
\end{thebibliography}
\end{document}